\documentclass[prd,twocolumn,superscriptaddress,altaffilletter,amssymb,showpacs,nofootinbib]{revtex4}
\usepackage[dvips]{graphicx}
\usepackage{amsmath}
\newcommand{\be}{\begin{equation}}
\newcommand{\ee}{\end{equation}}
\newcommand{\bea}{\begin{eqnarray}}
\newcommand{\eea}{\end{eqnarray}}

\begin{document}

\title{Evolution of density perturbations in double exponential quintessence models}

\author{Tame Gonzalez}\email{tame@uclv.edu.cu}
\author{Rolando Cardenas}\email{rcardenas@uclv.edu.cu}
\author{Israel Quiros}\email{israel@uclv.edu.cu}
\author{Yoelsy Leyva}\email{yoelsy@uclv.edu.cu}
\affiliation{Universidad Central de Las Villas, Santa Clara CP 54830, Cuba}

\date{\today}

\begin{abstract}
In this work we investigate the evolution of matter density
perturbations for quintessence models with a self-interaction
potential that is a combination of exponentials. One of the models
is based on the Einstein theory of gravity, while the other is
based on the Brans-Dicke scalar tensor theory. We constrain the
parameter space of the models using the determinations for the
growth rate of perturbations derived from data of the 2-degree
Field Galaxy Redshift Survey.
\end{abstract}


\maketitle

%%%%%%%%%%%%%%%%%%%%%%%%%%%%%%%%%%%%%%%%%%%%%%%%%%%%%%%%%%%%%%%%%%%%%%%%

\section{Introduction}

In the past few years, it has become apparent that the energy
budget of our universe is dominated by an unknown component called
"dark energy". The WMAP table of "best" cosmological parameters
\cite{lambda}, for instance, gives a $0.73\pm0.04$ abundance for
it, and a value of its equation of state $\omega<-0.78$. In order
to relieve some problems of the popular $\Lambda CDM$ model (like
the fine tuning issue), a dynamical $\Lambda$-term has been
proposed as representative of the dark energy. It's more popular
version is a slowly rolling scalar field named quintessence. Many
alternative cosmological models have been proposed, and indeed it
is a challenge the work of ruling out all the "incorrect ones" on
observational grounds. For instance many different potentials for
these self interacting scalar fields (quintessence) have been
proposed. However, it is obvious the importance of the
observational exploration of the proposed cosmological models, and
in this paper we give a further step in this direction.

A variety of quintessence self-interaction potentials  have been
studied. Among them, a single exponential is the simplest case.
This last model has two possible late-time attractors in the
presence of a barotropic fluid: a scaling regime where the scalar
field mimics the dynamics of the background fluid with a constant
ratio between both energy densities, or an attractor solution
dominated by the scalar field. Some of these models has been
studied in references \cite{cope,davi,ruba,Isra1}. Given that
single exponential potentials can lead to one of the above scaling
solutions, then it should follow that a combination of
exponentials should allow for a scenario where the universe can
evolve through a radiation-matter regime (attractor 1) and, at
some recent epoch, evolve into the scalar field dominated regime
(attractor 2). For this reason the combination of exponentials
represents an interesting alternative. Minimally coupled models
with double exponential potentials are studied in
\cite{ruba,Isra1}, meanwhile, Brans-Dicke (BD) models of
quintessence with this kind of potential have been studied, for
instance, in \cite{Isra1}.

The aim of this short paper is to (observationally) check models of
the universe with potentials that are combination of exponentials:
$V=V_1\exp(-\alpha \phi)+V_2\exp(-\beta \phi)$ (both in
minimally coupled (Einstein) and BD theories), by considering
another aspect of structure formation: the galaxy motions and
clustering, i. e., the evolution of density perturbations in the
Universe. In the past few years, observations of the large scale
structure of the Universe have improved greatly. The development of
fiber-fed spectrographs that can simultaneously measure spectra of
hundreds of galaxies has provided large redshift surveys such as the
2-degree Field Galaxy Redshift Survey (2dFGRS) and the Sloan Digital
Sky Survey (SDSS). In particular, the Anglo-Autralian Telescope of
the 2dFGRS has obtained the redshift of a quarter million galaxies.
This collaboration has produced abundant data and technical papers
\cite{2dFGRS} about galaxy motions and clustering, and we will refer
to some of this, in particular their velocity to density comparisons.

The paper is organized as follows: in section II we outline the
main characteristics of the models, in section III the main
aspects of velocity to density comparisons are exposed and the
equation for the growth of perturbations is solved, in section IV
the observational check is presented and interpreted, while in
section V conclusions are drawn.

\section{The Models}

In this section we supply the main equations characterizing the
dynamics of the models with double exponential potential of the
form:
\begin{equation}
V=V_{1}\;e^{-\alpha\phi}+V_{2}\;e^{-\beta\phi},\label{potential}
\end{equation}
where $V_1$, $V_2$, $\alpha$ and $\beta$ are free constant
parameters. We study separately models with minimal coupling
(basically Einstein gravity) and BD models written in Einstein frame
variables. We adopt also throughout the paper units with $8\pi
G=c=1$

\subsection{Einstein's Theory}

We analyze flat Friedmann-Robertson-Walker (FRW) solutions to
Einstein's theory with two fluids: a background fluid of ordinary
matter and a self-interacting scalar fluid that accounts for the
dark energy component.

If we set the following relationship between the barotropic index of the background ($\gamma$) and the free parameters of the above equation (\ref{potential}):
 
\begin{equation}\label{rel}
    \beta=\frac{3\gamma}{\alpha}
\end{equation}

then it can be show that
\begin{equation}
a(\tau)=\{\sqrt{\frac{1}{2-3\varepsilon}}\sinh[\mu_{E}(\tau+\tau_{0})]\}^{\frac{2}{3\gamma-\alpha^{2}}}.
\end{equation}
is an exact solution of the Einstein's field equations for the above
potential.\cite{Isra1}. Where

\begin{equation}\label{mu}
    \mu_{E}=(3\gamma-\alpha^{2})\sqrt{\frac{\gamma(2-3\varepsilon)}{3\gamma-\alpha^{2}}}
\end{equation}
and  $\tau_{0}$ is a constant of integration and the following time
coordinate $\tau$ has been introduced instead of the cosmic time
$t$: $dt=a^{\frac{\alpha^{2}}{2}}d\tau$.

This solution is very interesting since, the aforementioned
relationship between $\alpha$ and $\beta$ leads always to a
transition from a matter-dominated phase of the cosmic evolution at
high redshift, into a (late time) dark energy dominated phase. This
is, precisely, the kind of feature observational data suggests the
cosmic evolution should share. Other parameters of observational
interest are the Hubble expansion parameter:
\begin{equation}
H(\tau)=\sqrt{\frac{\gamma(2-3\varepsilon)}{3\gamma-\alpha^{2}}}a(\tau)^{-\frac{\alpha^{2}}{2}}\coth[\mu_{E}(\tau-\tau_{0})]
\end{equation}
the matter density parameter:
\begin{equation}
\Omega_{m}(\tau)=(1-\varepsilon)\left[\cosh[\mu_{E}(\tau-\tau_{0})]
\right]^{-2}
\end{equation}
and the equation of state (EOS) parameter:
\begin{equation}
\omega_{\phi}(\tau)=-1+\frac{\alpha^{2}}{3(1-\Omega_{m}(\tau))},
\end{equation}

This solution depends only on three free parameters ($\gamma,
\alpha, \varepsilon$), where $\varepsilon$ is the density
$\Omega_{\phi}(z)$ of dark energy in the early stages of the
evolution (high redshift $z\gg 1$). Nevertheless we will led with a
reduce parameters space($\alpha,\varepsilon$) because of we fix
$\gamma=1$, meaning cold dark matter dominance at present. Using
CMB\cite{CMB}, nucleosynthesis \cite{nucleo} and galaxy
formation\cite{galaxy} observations, the parameters space can be
constrained to be: $0<\alpha<1$ $\&$ $
0\leq\varepsilon\leq0.045$\cite{Isra1}.\footnote{Note that SN Ia is
not useful to constrain the free parameter $\alpha$, see
\cite{Isra1}. Indeed, it is well known the controversy about the
degeneracy of supernovae observations \cite{w,mb,mb2,w2,w3,mb3}.}

\subsection{Brans-Dicke gravity}

Now we study flat FRW exact solutions to BD theory with two
fluids: a background of ordinary matter and a self-interacting BD
scalar field fluid accounting for the dark energy in the universe.
In this case we are faced with two relevant frames (the Jordan
frame and the Einstein frame), in which BD theory can be
formulated. There has been discussion on whether these two frames
are equivalent \cite{equivalen}. It is not our aim to participate
in this controversy and for practical reasons (simplicity of
mathematical handling) we chose the Einstein frame.

In this frame, if one assumes the following
relationship\footnote{Recall that this relationship leads to a very
desirable pattern for the cosmic evolution. See similar discussion
under subsection A.} between the free parameters $\alpha$ and
$\beta$ in (\ref{potential}) and $\gamma$:

\begin{equation}
\beta=\frac{3\gamma}{\alpha}+\frac{4-3\gamma}{2\sqrt{\omega+\frac{3}{2}}},\label{BD}
\end{equation}
where $\omega$ is the BD coupling parameter; then
\begin{equation}
\bar{a}(r)=\{\sqrt{\frac{1}{2-3\varepsilon}}\sinh[\mu_{B}(r+r_{0})]\}^{\frac{2}{\alpha(\beta-\alpha)}}
\end{equation}
is an exact solution with:
\begin{equation}
    \mu_{B}=\frac{\alpha(\beta-\alpha)}{2}\sqrt{\frac{2-3\varepsilon}{3n^{2}(1-\varepsilon)}}
\end{equation}
and
\begin{equation}
    n=\frac{2\sqrt{\omega+\frac{3}{2}}-\alpha}{2\sqrt{\omega+\frac{3}{2}}}
\end{equation}

The time coordinate $r$ and the cosmic time in the Einstein frame
$\bar t$ are related by the following expression:
$d\bar{t}=\bar{a}^{\frac{\alpha^{2}}{2}}dr$. The bar notation means
we are working in the Einstein frame of BD theory. The other
interesting cosmological parameters are the Hubble expansion
parameter:

\begin{equation}\label{HBD}
\bar{H}(r)=\sqrt{\frac{2-3\varepsilon}{3n^{2}(1-\varepsilon)}}\bar{a}(\tau)^{-\frac{\alpha^{2}}{2}}\coth[\mu_{B}(r-r_{0})]
\end{equation}
the matter density parameter
\begin{equation}
\bar{\Omega}_{m}(r)=n^{2}(1-\varepsilon)\{\cosh[\mu_{B}(r+r_{0})]\}^{-2}
\end{equation}
and the equation of state parameter
\begin{equation}
\bar{\omega}_{\phi}(r)=-1+\frac{\alpha^{2}}{3(1-\bar{\Omega}_{m}(r))}
\end{equation}

This cosmological model also depends on three free parameters
($\gamma,\alpha,\varepsilon$). Like in the former case we fix the
value of $\gamma=1$ and executing a similar analysis with CMB,
nucleosynthesis and galaxy formation observations, the parameter
space can be constrained: $0<\alpha<0.37$ $\&$ $
0\leq\varepsilon\leq0.045$\cite{Isra1}.\footnote{Note again that SN
Ia observations are not useful to constrain the parameter space}.

\section{Perturbation growth}

We will perform here a preliminary study of the evolution of the
mass density contrast $(\delta=\delta\rho/\rho)$ in the mass
distribution, modelled as a pressureless fluid, in linear
perturbation theory. The evolution of quintessence density contrast
in not considered in this paper for simplicity. We understand,
however, that these perturbations might influence perturbations of
background matter\cite{mota}. This method is based on Newtonian
mechanics, that is better suited to the study of the development of
structure such as galaxies and clusters of galaxies. This
computation requires that we be able to isolate a region small
enough for the Newtonian gravitational potential energy and the
relative velocities within the region to be small (non
relativistic)\cite{peebles}. The equation of time evolution of mass
density contrast is

\begin{equation}\label{delt}
\ddot{\delta_{m}}+2H\dot{\delta_{m}}-4 \pi G \rho_{m} \delta_{m}=0,
\end{equation}
where the dot means derivative with respect to the comoving time.
In this equation the relative contribution of dark energy to the
energy budget enters into the expansion rate $H$. We shall
consider this equation in the matter dominated era, when the
radiation contribution is really negligible. The linear theory
relates the peculiar velocity field v and the density contrast by
\cite{peebles}

\begin{equation}
v(x)=H_{0} \frac{f}{4\pi} \int \delta_{m}(y) \frac{x-y}{|x-y|^3}
d^3y,
\end{equation}
where the growth index $f$ is defined as

\begin{equation}\label{Gindex}
f\equiv \frac{d\ln\delta_{m}}{d\ln a}.
\end{equation}
To solve the equation (\ref{delt}) for the evolution or
perturbations, it is useful to rewrite it in terms of suitable
variables, allowing some simplification

\begin{equation}\label{delta}
X^{2}(1+X^{2})\frac{d^{2}\delta_{m}}{dX^{2}}+X
\frac{d\delta_{m}}{dX}(X^{2}c+d)-e\delta_{m}=0,
\end{equation}
where the new variable $X$, the parameters $c$,$d$ and $e$ are
characteristic of model as follow:

\begin{table}
\caption{\label{tab:table1}This table show the characteristic
parameters for each model }
%\begin{ruledtabular}
\begin{tabular}{ccc}
parameters & Einstein's Theory &  BD gravity\\
\hline
    &  &  \\
  X & $\sinh[\mu_{E}(\tau+\tau_{0})]$ & $\sinh[\mu_{B}(r+r_{0})]$ \\
   &  &  \\
  c & $\frac{10-3\alpha^{2}}{2(3-\alpha^{2})}$ & $\frac{\alpha(\beta-2\alpha)+4}{\alpha(\beta-\alpha)}$ \\
   &  &  \\
  d & $\frac{4-\alpha^{2}}{2(3-\alpha^{2})}$ & $\frac{4-\alpha^{2}}{\alpha(\beta-\alpha)}$ \\
     &  &  \\
  e & $\frac{1}{2(3-\alpha^{2})}$ & $\frac{6n^{2}(1-\varepsilon)}{\alpha^{2}(\beta-\alpha)^{2}}$ \\
\end{tabular}
%\end{ruledtabular}
\end{table}

Equation (\ref{delta}) has two linearly independent solutions, the
growing mode $\delta_{m+}$ and the decreasing mode $\delta_{m-}$,
which can be expressed in terms of hypergeometric functions of
second type $_2 F_1$. We get

\begin{eqnarray}
  && \delta_{m+} \propto
X^{\frac{1}{2}(1-D+g)}._{2}F_{1}[\frac{1}{4}-\frac{d}{4}+\frac{1}{4}g
,-\frac{1}{4}+\frac{c}{2}-\nonumber\\ &&\frac{d}{4}+\frac{1}{4}g ,1+\frac{1}{2}g ,-X^{2} ]\nonumber
\end{eqnarray}
and
\begin{eqnarray}
  && \delta_{m-} \propto
X^{\frac{1}{2}(1-D-g)}._{2}F_{1}[\frac{1}{4}-\frac{d}{4}-\frac{1}{4}g
,-\frac{1}{4}+\frac{c}{2}-\nonumber\\&&\frac{d}{4}-\frac{1}{4}g ,1-\frac{1}{2}g ,-X^{2} ]\nonumber
\end{eqnarray}

where
\begin{equation}
    g=\sqrt{(d-1)^{2}+4e}.
\end{equation}
For $\tau \ll 1$ and $r\ll 1$ we can write

\begin{equation}\label{growthM}
\delta_{m+} \propto X^{\frac{1}{2}(1-D+g)},
\end{equation}
\begin{equation}\label{drecM}
    \delta_{m-} \propto
X^{\frac{1}{2}(1-D-g)}.
\end{equation}

For determining the growth index of the perturbations we use the
growing mode $\delta_{m+}$ (\ref{growthM}) and substitute into
(\ref{Gindex}). It is well known the biasing effect in galaxy
formation, i. e.; the relative perturbations in the galaxy field and
the matter field, on a point-by-point basis, are not equal:
\begin{equation}
\frac{\delta n}{n}(x)=b\frac{\delta \rho}{\rho}(x),
\end{equation}
where $n$ refers to the galaxy number density and $b$ is the bias
parameter. The parameter $\hat\beta=\frac{f}{b}$ relates the
growth rate $f$ of the perturbations (and hence the velocity field
of the galaxy motions) with the density bias $b$. In this sense
astrophysicists speak on a velocity/density comparison. Indeed, a
compelling agreement is seen to exist between the velocity and
density fields, which offers one possible test for the
gravitational instability picture for the origin of structure
\cite{ll}.

The 2dFGRS has measured the position and the redshift of a quarter
million galaxies and from the analysis of the correlation function,
determined the redshift distortion parameter $\hat\beta$ with the
bias parameter $b$ quantifying the difference between the galaxies
and the dark haloes distribution. Using the estimated $\hat\beta$
and the method employed by \cite{2dF1,2dF2,2dF3} to determine the
bias $b$, one may estimate $f=0.51\pm 0.11$ at the survey effective
depth $z=0.15$ . As we can see the growing mode depend of the free
parameters $(\alpha, \varepsilon)$ as the growth index $f$ to. Now
we use this fact to additionally constrain the above mentioned
parameter space.

\section{Observational Check}

\begin{figure}[t!]
\begin{center}
\hspace{0.4cm}\includegraphics[width=7.5cm,height=6cm]{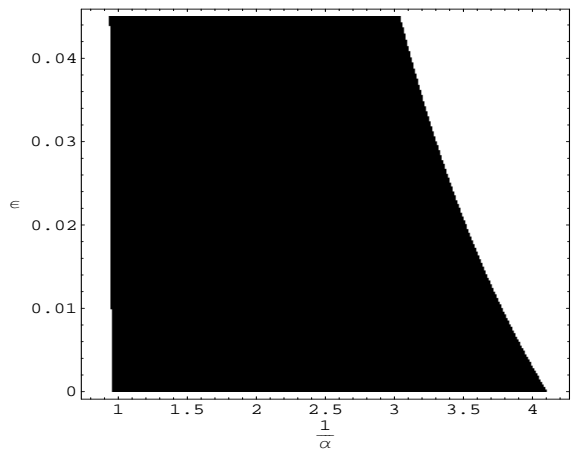}
\end{center}
\caption{Parameter space for the model based on Einstein gravity.
The parameter $\alpha$ is now constrained to a rather narrow region.}
\label{Efigure}
\end{figure}

Applying equation (\ref{Gindex}), in the flat FRW solutions to
Einstein's theory, does not give further constrain on $\varepsilon$,
but it does on $\alpha$, as Figure \ref{Efigure} show.

%\begin{figure}[t]
%\centering
%\leavevmode\epsfysize=5cm \epsfbox{Einstein.eps}\\
%\bigskip \caption[stabiA1]{\label{Efigure} Parameter space for the model based on Einstein gravity.The parameter $\alpha$ is now constrained to a rather narrow region.}
%\end{figure}

Assuming a flat universe, the parameter
$\varepsilon=\Omega_\varphi(\infty)$, is the amount of dark energy
in the very early universe ($z\sim\infty$). It is known that an
appreciable amount of dark energy at that epoch would imply an
expansion fast enough to prevent the formation of structure at
$z\sim3$, but this parameter had been already constrained in
Ref.\cite{Isra1}, so it is not problematic that now it has not been
additionally constrained. In this case, the velocity/density
comparison allows to locate $\alpha$ in a rather narrow region, thus
acting as a stringent selector of attractor-like solutions of the
kind \cite{cope,davi,ruba,Isra1}. Two things are to be said.
Firstly, Figure \ref{Efigure} shows the joint region ($\alpha, \varepsilon$)
satisfying the velocity/density comparison, for $\alpha$ alone we
might have a wider variation. Secondly, though narrow the interval,
still we can speak of $\alpha$ as a selector of a class of
solutions.

\begin{figure}[t!]
\begin{center}
\hspace{0.4cm}\includegraphics[width=7.5cm,height=6cm]{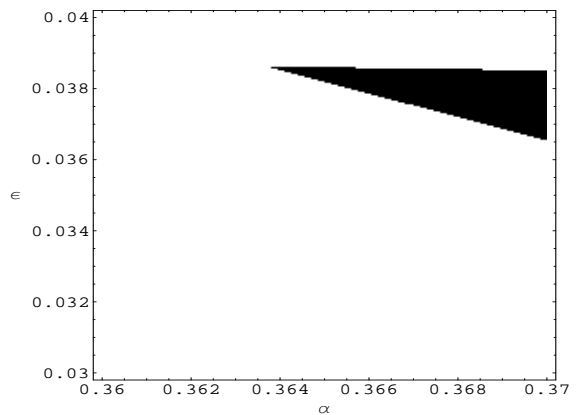}
\end{center}
\caption{Final parameter space for BD gravity. The shaded region obeys all of the
constrains considered here.}
\label{BD2}
\end{figure}

Figure \ref{BD2} shows the parameter space that was obtained for the model
based on Brans-Dicke gravity. Taking account this result and the
former obtained in \cite{Isra1} we construct the final parameter
space which obey all constrains considered so far.

%\begin{figure}[t]
%\centering
%\leavevmode\epsfysize=5cm \epsfbox{BD1.eps}\\
%\bigskip \caption[stabiA1]{\label{BD1} Parameter space for BD gravity obtained with the growth
%index analysis.}
%\end{figure}

%\begin{figure}[t]
%\centering
%\leavevmode\epsfysize=5cm \epsfbox{BD3.eps}\\
%\bigskip \caption[stabiA1]{\label{BD2} Final parameters space for BD gravity. This region obey all constrains(from \cite{Isra1} and growth index).}
%\end{figure}

\begin{figure}[t!]
\begin{center}
\hspace{0.4cm}\includegraphics[width=7.5cm,height=6cm]{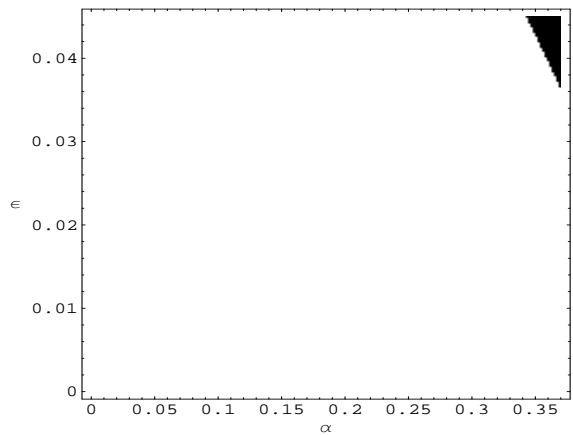}
\end{center}
\caption{Parameter space for BD gravity obtained with the growth
index analysis.}
\label{BD1}
\end{figure}

As we can see in Figure \ref{BD2} the parameters ($\alpha, \varepsilon$)
were already considerably constrained from the original
observations.

\section{Conclusions}

In the models presented here we found that the preliminary study
of perturbation growth is a good tool to constraint the parameter
space in quintessence models with a self-interaction potential
that is a combination of exponentials. This kind of potential
often appears in fundamental theories like the string or
supersymmetry. Besides, it can accommodate a pattern of cosmic
evolution that is characterized by a late time, dark energy
dominated attractor.

Research on the origins and evolution of the large-scale of the
universe is one of the hottest topics in cosmology. In this work,
we have used the relation between the peculiar velocity field of
the galaxies, the growth rate of perturbations and the density
bias in galaxy formation to make another step in the observational
check of two quintessence models, resulting in a further constrain
on the parameter space of the models. We consider that above
discussion adds another argument in favor of the use of
exponential potentials in quintessence cosmology. We plan in the
oncoming future proceed this works using another cosmological
probes, like CMB, for instance. It could be desirable to consider
perturbations of the quintessence field also.

\section{Acknowledgements}

We acknowledge the MES of Cuba by partial financial support of
this research. We are grateful also to David Wands for useful
conversations on the subject treated in this paper.

%%%%%%%%%%%%%%%%%%%%%%%%%%%%%%%%%%%%%%%%%%%%%%%%%%%%%%%%%%%%%%%%%%%%%%%%

\end{document}